\documentclass[%
 preprint,
superscriptaddress,
 amsmath,amssymb,
 aps,
prl,
]{revtex4-1}

\usepackage{graphicx}
\usepackage{dcolumn}
\usepackage{bm}
\usepackage[english]{babel} 
\usepackage{hyperref}
\usepackage{physics}
\usepackage{amsmath}
\usepackage{subfigure}

\begin{document}

\title{Comparing all-optical switching in synthetic-ferrimagnetic multilayers and alloys  }

\author{M. Beens}
\email[Corresponding author: ]{m.beens@tue.nl}
\affiliation{Department of Applied Physics, Eindhoven University of Technology \\ P.O. Box 513, 5600 MB Eindhoven, The Netherlands}
\author{M.L.M. Lalieu}
\affiliation{Department of Applied Physics, Eindhoven University of Technology \\ P.O. Box 513, 5600 MB Eindhoven, The Netherlands}
\author{A.J.M. Deenen}
\affiliation{Department of Applied Physics, Eindhoven University of Technology \\ P.O. Box 513, 5600 MB Eindhoven, The Netherlands}
\author{R.A. Duine }
\affiliation{Department of Applied Physics, Eindhoven University of Technology \\ P.O. Box 513, 5600 MB Eindhoven, The Netherlands}
\affiliation{ Institute for Theoretical Physics, Utrecht University \\
Leuvenlaan 4, 3584 CE Utrecht, The Netherlands}
\author{B. Koopmans}
\affiliation{Department of Applied Physics, Eindhoven University of Technology \\ P.O. Box 513, 5600 MB Eindhoven, The Netherlands}

\date{\today}

\begin{abstract}
We present an experimental and theoretical investigation of all-optical switching by single femtosecond laser pulses. Our experimental results demonstrate that, unlike rare earth-transition metal ferrimagnetic alloys, Pt/Co/[Ni/Co]$_N$/Gd can be switched in the absence of a magnetization compensation temperature, indicative for strikingly different switching conditions.  In order to understand the underlying mechanism, we model the laser-induced magnetization dynamics in Co/Gd bilayers and GdCo alloys on an equal footing, using an extension of the microscopic three-temperature model to multiple magnetic sublattices and including exchange scattering. In agreement with our experimental observations, the model shows that Co/Gd bilayers can be switched for a thickness of the Co layer far away from compensating the total Co and Gd magnetic moment.  We identify the switching mechanism in Co/Gd bilayers as a front of reversed Co magnetization that nucleates near the Co/Gd interface and propagates through the Co layer driven by exchange scattering. 
\end{abstract}

\maketitle

Femtosecond laser pulses provide a unique tool to manipulate magnetic order on ultrashort timescales. A prime example of this is all-optical switching (AOS) of magnetization by a single pulse, as first observed in ferrimagnetic GdFeCo alloys using circularly polarized laser pulses  \cite{Stanciu2007}. Later, AOS was demonstrated in the same material by using a single linearly polarized laser pulse \cite{Radu2011,Ostler2012}, which indicated that GdFeCo can be switched with ultrafast heating as the only stimulus.  

Deterministic AOS has a potential to be used in future magnetic memory devices, offering an ultrafast and energy-efficient way to write data.
Helicity-dependent AOS by circularly polarized pulses has been demonstrated in a wide variety of materials \cite{Hadri2016,Mangin2014,Medapalli2017}. However, in those cases the switch follows from a multiple-pulse mechanism. Purely thermal single-pulse AOS was  only observed in a limited number of materials systems, all including rare earth (Gd)-transition metal alloys \cite{Radu2011,Ostler2012,Guyader2016,Gorchon2017}. Very recently, it was also demonstrated for the  synthetic-ferrimagnetic layered structure \cite{Lalieu2017}, which allows for easy spintronic integration \cite{Lalieu2019}. The fact that single-pulse AOS is observed in both ferrimagnetic alloys and synthetic-ferrimagnetic multilayers raises the questions to what extent the switching of these materials systems relies on the same physics, and what  the specific conditions are for switching these materials systems.

In this work, we show that the conditions for single-pulse AOS in alloys (GdCo) and synthetic ferrimagnets (Co/Gd bilayers) are strikingly different. We experimentally demonstrate that single-pulse AOS in synthetic-ferrimagnetic Pt/FM/Gd is very robust, and can be achieved for a large range of the ferromagnetic (FM) layer thickness. The experiments indicate that the Pt/FM/Gd stacks can be switched in the absence of a compensation temperature.  In contrast, for alloys it is believed that it is crucial to have a compensation temperature near ambient temperature, such that the magnetization of the sublattices is compensated significantly  \cite{Guyader2016,Xu2017}. 

We performed simulations in order to understand this contrasting behaviour  and to identify the underlying mechanisms. The general theoretical framework for AOS describes the dynamics of multiple magnetic sublattices which are coupled antiferromagnetically. The intersublattice exchange coupling plays a crucial role, transferring angular momentum between the sublattices  \cite{Mentink2012}. Different approaches have been made to describe the spin dynamics of the magnetic sublattices, e.g., the atomistic Landau-Lifshitz-Gilbert equation  \cite{Ostler2011,Atxitia2012,Barker2013,Gerlach2017,Atxitia2018} and the microscopic three-temperature model (M3TM) \cite{Koopmans2010,Schellekens2013}. Here, we use the latter microscopic description, in which it is assumed that angular momentum transfer between the sublattices is mediated by exchange scattering \cite{Schellekens2013}.  We derive an analytical expression for the magnetization dynamics resulting from the exchange scattering between (i) the sublattices in a GdCo alloy and (ii) the atomic monolayers in a Co/Gd bilayer. The model reproduces the distinct role of the compensation temperature. Moreover, it shows that the robustness of AOS in the Co/Gd bilayers can be explained by the non-local character of the switching mechanism, which we identify as a front of reversed Co magnetization that propagates away from the interface.

This work starts with a brief description of the experimental methods and results. After that,  the theoretical framework will be introduced. For the sake of direct comparison, we focus our theoretical discussion on the magnetization dynamics in GdCo alloys and Co/Gd bilayers.  We present phase diagrams that show qualitatively the switching conditions and point out the differences for both materials systems. Finally, the typical switching mechanism of the bilayers is explained explicitly.

The experiments are performed using Si:B(substrate)/Ta(4)/Pt(4)/FM/Gd(3)/Pt(2) stacks (thickness in nm), which are deposited at room temperature using dc magnetron sputtering at $10^{-8} \mbox{ mbar}$ base pressure. In this work, Co(0.2)/[Ni(0.6)/Co(0.2)]$_N$ multilayers are used for the FM layer, with $N$ repeats ranging from $N = 2$ to $5$. Using polar magneto-optical Kerr effect measurements, a square hysteresis loop with 100\% remanence was obtained for all samples, confirming the presence of a well-defined perpendicular magnetic anisotropy in the samples. 

\begin{figure}[t!]
\includegraphics[scale=0.96]{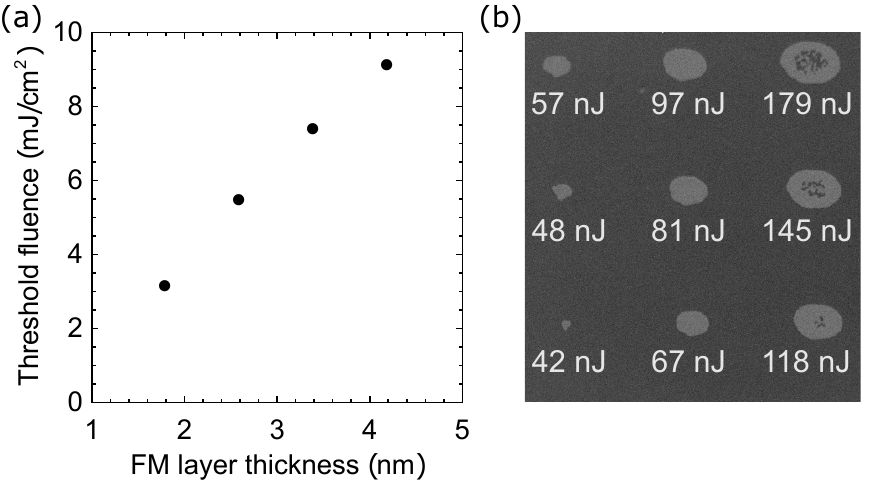}
\caption{\label{fig:figure1} (a) Threshold fluence as a function of the FM layer thickness in a Pt/FM/Gd stack. The black dots are measured using a FM layer composed of Co(0.2)/[Ni(0.6)/Co(0.2)]$_N$ multilayers for $N=2,3,4,5$. The error margins are small compared to the scale of the figure. (b) A Kerr microscope image of the (initially saturated) Co/Ni sample with $N=3$ after excitation with single linearly polarized laser pulses with different pulse energies. 
} 
\end{figure}

The response of the magnetization in the Pt/FM/Gd stacks to laser-pulse excitation was investigated using linearly polarized laser pulses with a central wavelength of $700 \mbox{ nm}$ and a pulse duration of $\approx 100\mbox{ fs}$. The measurements are performed at room temperature, and start by saturating the magnetization using an externally applied field. Then, the external field is turned off, and the sample is exposed to single laser pulses with varying pulse energies. The response of the magnetization to the laser-pulse excitation is measured in the steady state (i.e., long after the excitation) using a magneto-optical Kerr microscope. 

A typical result of the AOS measurement for the sample with $N = 3$ is presented in Fig. \ref{fig:figure1}(b). The figure displays the Kerr image of the (initially saturated, dark) sample after excitation with single linearly polarized laser pulses with different pulse energies. The figure shows clear homogeneous domains with an opposite magnetization direction (light) being written by the laser pulses. Moreover, the domain size increases for increasing pulse energy, as is expected when using a Gaussian pulse shape. For the highest pulse energies a multidomain state is formed in the centre region of the domain, where the lattice is heated above the Curie temperature \cite{Gorchon2016}. 

The AOS-written domain size as a function of the pulse energy can be used to determine the threshold fluence 
\cite{Lalieu2017,Liu1982}. Figure \ref{fig:figure1}(a) displays the threshold fluence as a function of the (total) thickness of the Co(0.2)/[Ni(0.6)/Co(0.2)]$_N$ multilayer. The results show that decreasing the thickness of the FM layer leads to a lower threshold fluence. This behaviour is reproduced in the model calculations that are presented later. It can be partially explained by a decrease in the Curie temperature with film thickness in the thin-film limit, but it will be shown that also other processes are involved. 

Remarkably, single-pulse AOS is seen for up to 5 repeats, corresponding to a FM layer thickness of $4.2 \mbox{ nm}$. For these relatively thick FM layers, the total magnetic moment of the FM layer is much larger than the induced magnetic moment in the Gd layer corresponding to approximately $1$-$2$ atomic monolayers of fully saturated Gd  \cite{Lalieu2017}, i.e., the system is far from compensated. Hence, the experiment indicates that the switching mechanism in the bilayers is independent of a possible compensation temperature. To understand the underlying mechanism, we developed a simplified model.

Analogous to Schellekens et al. \cite{Schellekens2013}, we assume that separate spin subsystems are coupled to a single electron and phonon subsystem.  Like in the basic M3TM \cite{Koopmans2010}, the electrons are treated as a spinless free electron gas and the phonons are described within the Debye model. It is assumed that both subsystems are internally thermalized, and that the electron temperature $T_e$ and phonon temperature $T_p$ are homogeneous. The spin specific heat is neglected. Femtosecond laser heating is modelled by adding an energy source to the electron subsystem. Heat diffusion to the substrate is added to the phonon subsystem as an energy dissipation term with timescale $\tau_D$. The spin subsystems, labeled with index $i$, are treated within a Weiss mean field approach. At each lattice site $D_{\mathrm{s},i} = \mu_{\mathrm{at},i}/2 S_i$ spins are present, where $\mu_{\mathrm{at},i}$ is the atomic magnetic moment (in units of the Bohr magneton $\mu_B$) and $S_i$ is the spin quantum number. 

\begin{figure*}[ht!]
\includegraphics[scale=0.86]{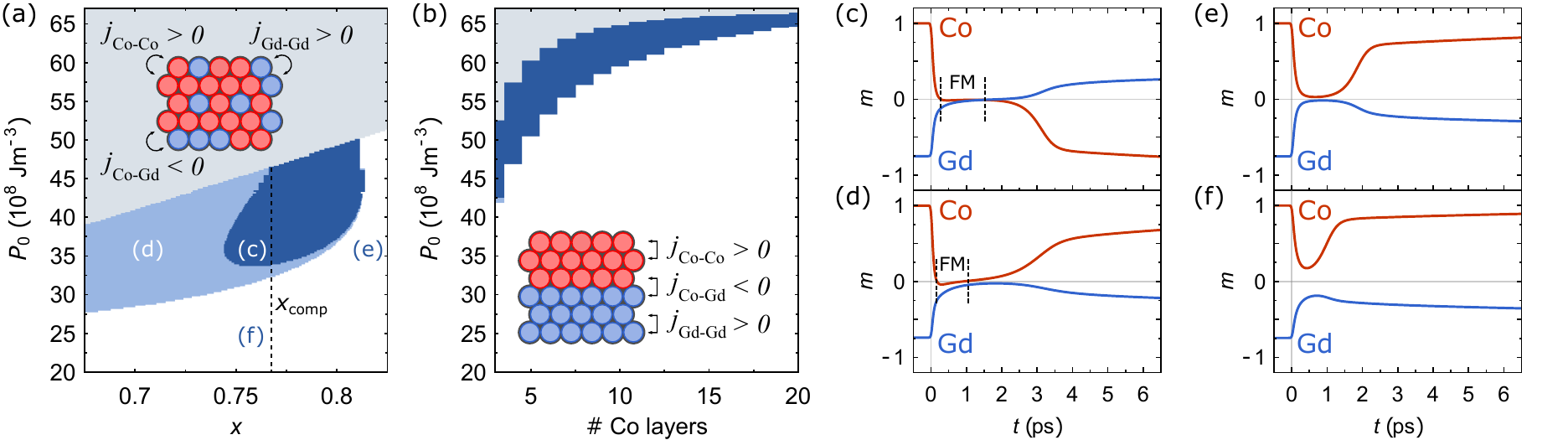}
\caption{\label{fig:figure2} Phase diagram for AOS as a function of the laser pulse energy $P_0$ and (a) $\mbox{Co}$ concentration $x$ for a $\mbox{Gd}_{1-x} \mbox{Co}_x$ alloy, (b) the number of Co monolayers in a Co/Gd bilayer. The dark blue regions indicate a switch in the final state (c) and the white regions indicate  no switch (e)-(f). Light blue indicates a transient ferromagnetic state, but no switch (d). The grey regions indicate  that the  phonon temperature $T_p$ exceeds the Curie temperature $T_C$.  The dashed line in figure (a) indicates the Co concentration $x_{\mathrm{comp}}$ for which the compensation temperature is equal to room temperature. The insets in (a) and (b) schematically show the modelled system, including the exchange parameters. Figure (c)-(f) display the element-specific magnetization dynamics in $\mbox{Gd}_{1-x}\mbox{Co}_x$ for different values for $x$ and $P_0$, corresponding to the various regions in (a). 
} 
\end{figure*}

For the $\mbox{Gd}_{1-x} \mbox{Co}_x $ alloys, we define a normalized magnetization $m_i$ for each of the two sublattices. As depicted in the inset of Fig. \ref{fig:figure2}(a), the exchange field experienced by each atom depends on the type of atom and the composition of its nearest neighbours. Hence, the exchange splitting is given by

\begin{eqnarray}
\Delta_{\mathrm{Co}} &=& x \gamma_{\mathrm{Co}\mbox{-}\mathrm{Co}} m_{\mathrm{Co}} +(1-x) \gamma_{\mathrm{Co}\mbox{-}\mathrm{Gd}} m_{\mathrm{Gd}},
\\
\Delta_{\mathrm{Gd}} &=& x \gamma_{\mathrm{Gd}\mbox{-}\mathrm{Co}} m_{\mathrm{Co}} +(1-x) \gamma_{\mathrm{Gd}\mbox{-}\mathrm{Gd}} m_{\mathrm{Gd}},
\end{eqnarray}

\noindent 
where we defined $\gamma_{ij}= j_{ij} z D_{\mathrm{s},j} S_j$ ($i,j \in \{\mbox{Co,Gd}\}$) in terms of the (intra- or intersublattice) exchange coupling constant $j_{ij}$ and the number of nearest neighbours $z$. Note that $j_{\mathrm{Co}\mbox{-}\mathrm{Gd}}$ is negative and quantifies the strength of the antiferromagnetic coupling between the Co and Gd sublattices. 

For the Co/Gd bilayers we introduce a normalized magnetization $m_i$ for each atomic monolayer $i$ separately. Each layer only interacts with its adjacent layers. For simplicity, we assume that the separate layers lie in the $(111)$ plane of an fcc lattice. This means that each atom has 6 nearest neighbours in the same layer and 3 nearest neighbours in each adjacent layer. Thus, the exchange splitting of layer $i$ is 

\begin{eqnarray}
\Delta_i &=& \dfrac{\gamma_{i,i-1}}{4} m_{i-1}  + \dfrac{\gamma_{i,i}}{2} m_i + \dfrac{\gamma_{i,i+1}}{4} m_{i+1}.
\end{eqnarray}

\noindent Note that the antiferromagnetic coupling, proportional to $j_{\mathrm{Co}\mbox{-}\mathrm{Gd}}$, is only experienced by the layers adjacent to the interface (see inset Fig. \ref{fig:figure2}(b)). 

We include two channels for angular momentum transfer. Elliott-Yafet spin-flip scattering mediates the transfer of angular momentum between the spin subsystems and the lattice \cite{Koopmans2005}. An extension of the  M3TM, which accounts for spin systems with arbitrary spin $S$, is derived to describe the resulting magnetization dynamics \cite{Koopmans2010,Cywinski2007,Supplementary}. Here, we take $S_{\mathrm{Co}}=1/2$ and $S_{\mathrm{Gd}}=7/2$, for which the Weiss model is well fitted to the experimental data for the magnetization as a function of temperature \cite{Cullity1972,Coey2009}. Angular momentum transfer between the different spin subsystems is mediated by exchange scattering \cite{Schellekens2013}. In this \textit{e-e} scattering process, spins originating from different subsystems are flipped in the opposite direction.  We use Fermi's golden rule to find an analytical expression for the magnetization dynamics resulting from the exchange scattering (see Supplemental Material Section II) \cite{Supplementary}. For $i\neq j$ we have 

\begin{widetext}
\begin{eqnarray}
\label{eq:fermi} 
\dfrac{dm_i}{dt}\bigg|_{\mbox{ex}} &=& \dfrac{2 \eta_{ij} C_j}{\mu_{\mathrm{at},i}} T_e^3
 \bigg[\, \sum_{s=-S_i+1\,\,}^{S_i } \sum_{\,\,s'=-S_j}^{S_j-1}  W^{-+}_{ij;ss'}(\Delta_{i}-\Delta_j )f_{i,s} f_{j,s'}
 \\ && \,\qquad\qquad 
 \nonumber   
 -\sum_{s=-S_i\,\,}^{S_i-1 } \sum_{\,\,s'=-S_j+1}^{S_j } W^{+-}_{ij;ss'}(\Delta_{i}-\Delta_j )f_{i,s} f_{j,s'} 
 \bigg].
\end{eqnarray}
\end{widetext}

\noindent 
The indices $s$ and $s'$  correspond to the $z$ component of the spin and label the discrete energy levels. The average occupation of level $s$ in spin subsystem $i$ is given by $f_{i,s}$, and  $ \Delta_i-\Delta_j$ is the energy difference between the initial and final spin configuration.  The dimensionless function $W^{\pm\mp}_{ij;s s'}  $ parametrizes the transition rate from level $s$ to $s\pm 1$ in subsystem $i$ and level $s'$ to $s'\mp 1$ in subsystem $j$. The coordination number $C_j$ counts the relative number of nearest neighbours that are part of spin subsystem  $j$. For alloys $C_j$ is given by $C_{\mathrm{Co}}=12 x$ and $C_{\mathrm{Gd}}=12(1-x)$. For bilayers we have $C_j=3$, the number of nearest neighbours in an adjacent layer. The constant $\eta_{ij}$ is determined by the matrix element of the exchange scattering Hamiltonian \cite{Supplementary}, for which we assume that it is proportional to the exchange coupling constant. Hence, we write $\eta_{ij} \propto \lambda_{ij} j_{ij}^2 $ , where $\lambda_{ij}$ is a dimensionless parameter. In the following discussions we assume that $\lambda_{ij}= \lambda = 5$, which is chosen in order to retrieve realistic results from the simulations (e.g., for $\lambda=1$ no switching is found). Results for different choices of $\lambda$ and $S_{\mathrm{Gd}}$ are presented in the Supplemental Material \cite{Supplementary}. Note that for the bilayers, equation (\ref{eq:fermi}) should include terms for the interaction with both adjacent layers, i.e., $j=i+1$ and $j=i-1$, and the full expression is given by the sum of these two terms. 

The temporal profile of the laser pulse is modelled by a Gaussian function $P(t) = (P_0/(\sigma\sqrt{\pi})) \mbox{Exp}(-(t-t_0)^2/\sigma^2)$, where $P_0$ is the absorbed laser pulse energy density and $\sigma $ is the pulse duration, which is set to $ 50\mbox{ fs}$. We assume that the laser pulse heats up the system homogeneously, which is a valid approximation for the systems we model, e.g. Co/Gd bilayers containing up to 20 Co atomic monolayers. We note that for thicker systems the approximation becomes questionable, and a finite penetration depth should be incorporated into the modelling. 

The laser-induced dynamics of $m_i(t)$ is calculated numerically. We assume that the spin subsystems are not necessarily in internal equilibrium, meaning that after excitation the ratio between $f_{i,s}$ and $f_{i,s\pm 1}$ is not given by a Boltzmann distribution, and we need to solve a set of $2S_i+1$ coupled differential equations for each spin subsystem $i$ \cite{Supplementary}.  The exact values for the material parameters, including the exchange coupling constants $j_{ij}$, are listed in the Supplemental Material   \cite{Supplementary}.

Two phase diagrams are constructed that display the occurrence of AOS as a function of the laser pulse energy $P_0$ and (i) the Co concentration $x$ of a $\mbox{Gd}_{1-x}\mbox{Co}_x$ alloy, (ii) the number of Co monolayers for a $\mbox{Co}/\mbox{Gd}$ bilayer (the Co thickness). We assume that the ambient temperature is equal to room temperature ($T_\mathrm{amb}=295\mbox{ K}$). The result is shown in Fig. \ref{fig:figure2}(a) and Fig. \ref{fig:figure2}(b). The color scheme indicates  whether the magnetization of the Co is reversed after relaxation, which is determined by calculating its sign at $t=100\mbox{ ps}$. For the bilayers we take the average of the magnetization of the Co monolayers. Figure \ref{fig:figure2}(c)-(f) are presented to clarify the meaning of the color scheme, and show the corresponding element-specific magnetization dynamics for the alloys. In the phase diagrams, the dark blue regions indicate that the Co magnetization is reversed, meaning that AOS has occurred (Fig. \ref{fig:figure2}(c)). The light blue regions indicate that there is a transient ferromagnetic state created, but after relaxation the magnetization is switched back to its initial direction (Fig. \ref{fig:figure2}(d)). The white regions indicate that the magnetization relaxes to its initial direction, without a transient ferromagnetic state (Fig. \ref{fig:figure2}(e)-(f)).  The grey regions indicate that the maximum of the phonon temperature $T_p$ exceeds the Curie temperature. In the experiments, this would likely result in the creation of a multidomain state \cite{Gorchon2016}. 

The vertical dashed line in Fig. \ref{fig:figure2}(a) indicates the compensation point $x_{\mathrm{comp}}\sim 0.77$, the Co concentration for which the total magnetic moment of the alloy is zero at room temperature. The dark blue region shows that the alloys can only be switched in a limited range of the Co concentration, sufficiently close to the compensation point. Furthermore, the minimum threshold fluence is found to be close to the compensation point. These findings are in agreement with the experiments \cite{Xu2017}. From the phase diagram we can conclude that in order to switch the alloy, a significant magnetization compensation is necessary. Hence, the model yields that the magnetization compensation temperature plays a crucial role in switching the alloys. 

\begin{figure}[t!]
\includegraphics[scale=0.95]{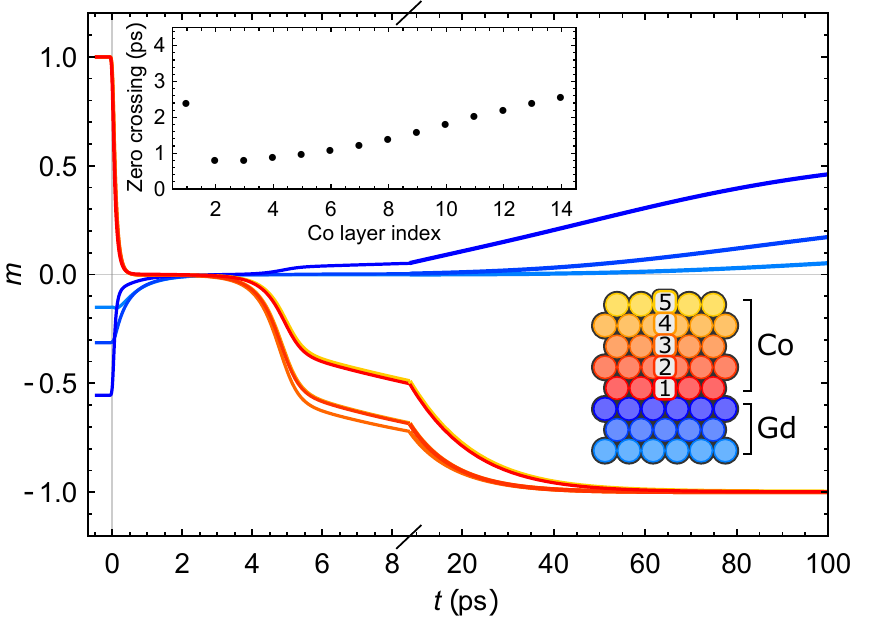}
\caption{\label{fig:figure3} Laser-induced  magnetization dynamics of all atomic monolayers in a Co/Gd bilayer consisting of 5 Co monolayers and 3 Gd monolayers for $P_0=55\cdot 10^8 \mbox{ Jm}^{-3}$. The inset shows the time at which the magnetization of each Co monolayer is reversed for a system of 14 Co layers and 3 Gd layers for $P_0=65\cdot 10^8 \mbox{ Jm}^{-3}$ (index $1$ corresponds to the Co layer adjacent to the interface).  } 
\end{figure}

A clear  difference is found when we compare this to the situation for bilayers,  Fig. \ref{fig:figure2}(b). This phase diagram shows that the bilayers can be switched for a relatively large number of Co monolayers, even though the threshold fluence increases as a function of the number of Co monolayers. More specifically, even bilayers with 20 Co monolayers can be switched. For these bilayers, the ratio of the total Co and Gd magnetic moment is $\mu_{\mathrm{Co}}/\mu_{\mathrm{Gd}}\sim 4$ (at $T_\mathrm{amb}=295\mbox{ K}$), which is significantly far from compensation ($\mu_{\mathrm{Co}}/\mu_{\mathrm{Gd}}=1$). In contrast, for the alloys switching only occurs in the range $\mu_{\mathrm{Co}}/\mu_{\mathrm{Gd}}\sim 0.9\mbox{ - }1.3$. Note that for convenience we described the FM layer as pure Co, whereas in the experiments Co/Ni multilayers are used. Including the Co/Ni multilayers will not change the qualitative properties of the switching mechanism.  Hence, the model agrees well with our experimental observation that shows single-pulse AOS in Pt/FM/Gd for relatively thick FM layers, and verifies that the magnetization compensation temperature does not play a crucial role in switching the synthetic ferrimagnets.

A more detailed analysis of the typical switching mechanism in the bilayers is presented in Fig. \ref{fig:figure3}, which shows AOS in a system of 5 Co monolayers and 3 Gd monolayers. We plotted the normalized magnetization of the separate layers as a function of time after laser pulse excitation (at $t=0$). The inset displays the time at which each Co monolayer reverses its magnetization direction, for a system of 14 Co layers and 3 Gd layers. The inset clearly shows that the Co layers are switched consecutively, starting with the Co layers near the Co/Gd interface. Triggered by the laser pulse, the switch is initiated near the interface due to exchange scattering between the adjacent Co and Gd monolayers. The dynamics of the first Co monolayer (Index 1 in Fig. \ref{fig:figure3}) is strongly modified by the exchange field from the slowly demagnetizing Gd layer \cite{Supplementary}. Hence, the second Co monolayer is switched first. Subsequently, the switch propagates throughout the Co layer driven by exchange scattering between neighbouring Co monolayers. This successive switching mechanism, with a front of reversed Co magnetization propagating away from the interface, can succeed independently of the number of Co monolayers and explains why the Co/Gd bilayer can be switched for a relatively large Co thickness. 

To conclude, both the experiment and the theoretical model show that single-pulse AOS switching in synthetic-ferrimagnetic bilayers is independent of a possible compensation temperature, whereas in ferrimagnetic alloys the compensation temperature plays a crucial role. We identified the propagation of a switching front as the characteristic  mechanism for AOS in the bilayers. These new insights show that single-pulse AOS in synthetic ferrimagnets is more robust than  in  ferrimagnetic alloys, and emphasize that Pt/FM/Gd synthetic ferrimagnets are a very promising candidate for integration of single-pulse AOS in future data storage devices.  

This work is part of the research programme of the Foundation for Fundamental Research on Matter (FOM), which is part of the Netherlands Organisation for Scientific Research (NWO).

\providecommand{\noopsort}[1]{}\providecommand{\singleletter}[1]{#1}%

\end{document}


\title{Supplemental Material for: ``Comparing all-optical switching in synthetic-ferrimagnetic multilayers and alloys'' }

\author{M. Beens}
\email[Corresponding author: ]{m.beens@tue.nl}
\affiliation{Department of Applied Physics, Eindhoven University of Technology \\ P.O. Box 513, 5600 MB Eindhoven, The Netherlands}
\author{M.L.M. Lalieu}
\affiliation{Department of Applied Physics, Eindhoven University of Technology \\ P.O. Box 513, 5600 MB Eindhoven, The Netherlands}
\author{A.J.M. Deenen}
\affiliation{Department of Applied Physics, Eindhoven University of Technology \\ P.O. Box 513, 5600 MB Eindhoven, The Netherlands}
\author{R.A. Duine }
\affiliation{Department of Applied Physics, Eindhoven University of Technology \\ P.O. Box 513, 5600 MB Eindhoven, The Netherlands}
\affiliation{ Institute for Theoretical Physics, Utrecht University \\
Leuvenlaan 4, 3584 CE Utrecht, The Netherlands}
\author{B. Koopmans}
\affiliation{Department of Applied Physics, Eindhoven University of Technology \\ P.O. Box 513, 5600 MB Eindhoven, The Netherlands}

\date{\today}

\maketitle

\onecolumngrid

\section{I.$\qquad$ M3TM for systems with an arbitrary spin}
\renewcommand{\theequation}{S.\arabic{equation}}
\renewcommand{\thefigure}{S.\arabic{figure}}
\renewcommand{\thetable}{S.\arabic{table}}

\noindent We consider the magnetization dynamics resulting from Elliott-Yafet spin-flip scattering. The derivation is analogous to the derivation of the M3TM \cite{Koopmans2010}, except that we describe spin systems with an arbitrary spin quantum number S. In this derivation we only display the extra steps that are needed to extend the M3TM to arbitrary spin. The skipped steps are fully equivalent to the derivation for $S=1/2$ \cite{Koopmans2010}. 

We consider a spin system with an arbitrary spin S, which is treated within a Weiss mean field approach. Each spin corresponds to a system of $2S+1$ energy levels, where the energy is determined by the $z$ component of the spin. The energy difference between the spin levels is given by the exchange splitting $\Delta$. The average occupation of each level  is given by $f_s$, where $s$ labels the $z$ component of the spin. We write \cite{Cywinski2007}

\begin{eqnarray}
\label{eq:sup1}
\dfrac{dm}{dt } &=& -\dfrac{1}{S} \sum_{s=-S}^{S} s \dfrac{d f_{s}}{dt} .
\end{eqnarray}

\noindent The dynamics of $f_s$ can be determined by calculating the transition rates for the states $s\rightarrow s\pm 1$. In the presence of Elliott-Yafet spin-flip scattering, we write

\begin{eqnarray}
\label{eq:sup2} 
\dfrac{df_s }{dt}\bigg|_{\mbox{EY} } &=& -[W^-_{\mathrm{EY},s}(\Delta) +W^+_{\mathrm{EY},s}(\Delta) ]f_s 
+ W^-_{\mathrm{EY},s+1}(\Delta) f_{s+1} + W^+_{\mathrm{EY},s-1}(\Delta) f_{s-1} ,
\end{eqnarray}

\noindent where $\Delta$ corresponds to the exchange splitting and $W_{\mathrm{EY},s}^{\pm}(\Delta)$ parametrizes the transition rates for each allowed transition \cite{Cywinski2007}. By applying Fermi's golden rule, using the same approximations and simplifications as the derivation of the M3TM \cite{Koopmans2010}, we find

\begin{eqnarray}
W_{\mathrm{EY},s}^{\pm}(\Delta) &=&  R\cdot  \dfrac{ T_p}{T_C}\cdot  \dfrac{\Delta}{2k_B T_C}  \cdot \dfrac{ e^{\mp\Delta/2k_B T_e } }{2 \mbox{ sinh}(\Delta/2k_B T_e )} (S(S+1)-s(s\pm 1)) .
\end{eqnarray}


\noindent The last term corresponds to the squared eigenvalue of the spin ladder operator $\hat{S}^{\pm} $. The scattering rate is proportional to the material constant $R$ given in the units $\mbox{ps}^{-1}$, which will be discussed below. It can easily be shown that for $S=1/2$ this expression results in the basic M3TM model \cite{Koopmans2010}

\begin{eqnarray}
\dfrac{dm}{dt}\Big|_{\mathrm{EY} }  &=& R m \dfrac{ T_p}{T_C}  \Big(1-m \mbox{coth}\Big(\dfrac{m T_C}{ T_e}  \Big) \Big) .
\end{eqnarray}


\noindent In our Letter, we use the basic M3TM  to describe the dynamics resulting from Elliott-Yafet scattering in Co ($S_{\mathrm{Co}}=1/2$). In contrast, for Gd  we choose $S_{\mathrm{Gd}}=7/2 $ and equation \ref{eq:sup2} needs to be solved for all $f_{\mathrm{Gd},s}$. This choice is made because for $S_\mathrm{Gd}=7/2$ the Weiss model is well fitted to the experimental data for the magnetization as a function of temperature \cite{Coey2009}. However, it needs to be noted that this choice is not trivial, since the driving spin-flip processes are dominated by  Elliott-Yafet processes in the $5d6sp$ valence band \cite{Koopmans2010}. For the latter, the choice $S=1/2$ would be more convenient. From our simulations we know that the demagnetization rate in an isolated system (e.g. pure Gd) is mainly determined by the value for $R$ and $T_C$, and that the effect of a different value for the spin quantum number $S$ is relatively unimportant. However, the spin quantum number $S$ has a significant influence on the dynamics resulting from the exchange scattering and the choice for $S_{\mathrm{Gd}}=7/2$ leads to significantly better results compared to the all-optical switching experiments, as will be discussed below. Therefore, we take $S_{\mathrm{Gd}}=7/2$ and we use the value for $R_{\mathrm{Gd}}$ reported in \cite{Koopmans2010}. 
 

\section{II.$\qquad$Details about exchange scattering} 

\noindent To derive an expression for the magnetization dynamics resulting from the exchange scattering we start with the Hamiltonian \cite{Schellekens2013}

\begin{eqnarray}
\label{eq:hex} 
\hat{H}_{\mathrm{ex},ij} = \sum_{\mathbf{k}} \sum_{\mathbf{k}'}  \sum_{\mathbf{k}''}  \sum_{\mathbf{k}'''}  \sum_v^{ND_{\mathrm{s},i} } \sum_w^{ND_{\mathrm{s},j} } 
\Big(\dfrac{j_{\mathrm{ex},ij}}{N^2}\Big) 
c_{\mathbf{k}'''}^\dagger c_{\mathbf{k}''}^\dagger c_{\mathbf{k}'} c_{\mathbf{k}} 
(\hat{S}^+_{i,v}\hat{S}^-_{j,w} + \hat{S}^-_{i,v}\hat{S}^+_{j,w} ). 
\end{eqnarray}

\noindent This Hamiltonian describes the processes where an electron-electron scattering event results in a change of the z component of a spin from subsystem $i$ at lattice point $v$ and a spin from subsystem $j$ at lattice point $w$.  We introduce the function $\Gamma^{\pm \mp}_{ij, s s'}$, which corresponds to the rate of the transitions where $S_{z,i,v}=s\rightarrow s\pm 1$   and $S_{z,j,w}=s'\rightarrow s'\mp 1$. In the following, we only take into account nearest neighbour interactions and we assume that the spin systems are homogeneous (independent of lattice points $v$ and $w$). By applying Fermi's golden rule we write 

\begin{eqnarray}
\Gamma^{\pm\mp}_{ij, s s'}  &=& \dfrac{2\pi}{\hbar} C_j D_{\mathrm{s},j}   \sum_{\mathbf{k}} \sum_{\mathbf{k}'}  \sum_{\mathbf{k}''}  \sum_{\mathbf{k}'''}   
\Big(\dfrac{j_{\mathrm{ex},i j}}{N^2}\Big)^2 f_{\mathbf{k}} f_{\mathbf{k}'}(1-f_{\mathbf{k}''})(1-f_{\mathbf{k}'''})
\delta(\epsilon_{\mathbf{k}}+\epsilon_{\mathbf{k}'} \mp \Delta_{ij}-\epsilon_{\mathbf{k}''}-\epsilon_{\mathbf{k}'''} ) S_{ij,s s'}^{\pm\mp}, 
\end{eqnarray}

\noindent where we defined $\Delta_{ij} = \Delta_i-\Delta_j $ and $C_j$ is the coordination number, which is discussed in the main text. $f_{\mathbf{k}}$ corresponds to the distribution function for the electrons and $\epsilon_{\mathbf{k}}$ corresponds to the single-particle energy of an electron with momentum $\mathbf{k}$.  $S_{ij,s s'}^{\pm\mp}$ contains the squared eigenvalues of the spin ladder operators given by

\begin{eqnarray}
S^{\pm\mp}_{ij,ss'} &=& (S_i(S_i+1)-s(s\pm 1)) (S_j(S_j+1)-s'(s'\mp 1)) .
\end{eqnarray}

\noindent We take the continuum limit to carry out the summations over  $\mathbf{k}$. Then, we write

\begin{eqnarray}
\Gamma^{\pm\mp}_{ij, s s'}  &=& \dfrac{\eta_{ij} C_j T_e^3}{D_{\mathrm{s},i}}  \mbox{SI}(\mp\Delta_{ij} /k_B T_e)  S_{ij,ss'}^{\pm\mp}.
\end{eqnarray} 

\noindent The scattering integral $\mbox{SI}(\mp\Delta_{ij}/k_B T_e ) $ parametrizes the probability of an electron-electron scattering event to occur and is given by

\begin{eqnarray}
\mbox{SI}(\pm \Delta x) \equiv  
\int d x_1 \int dx_2 \int dx_3
(1-f(x_1+x_2 \pm \Delta x -x_3))(1-f(x_3))f(x_2)f(x_1),
\end{eqnarray}

\begin{figure}[t!]
\includegraphics[scale=0.85]{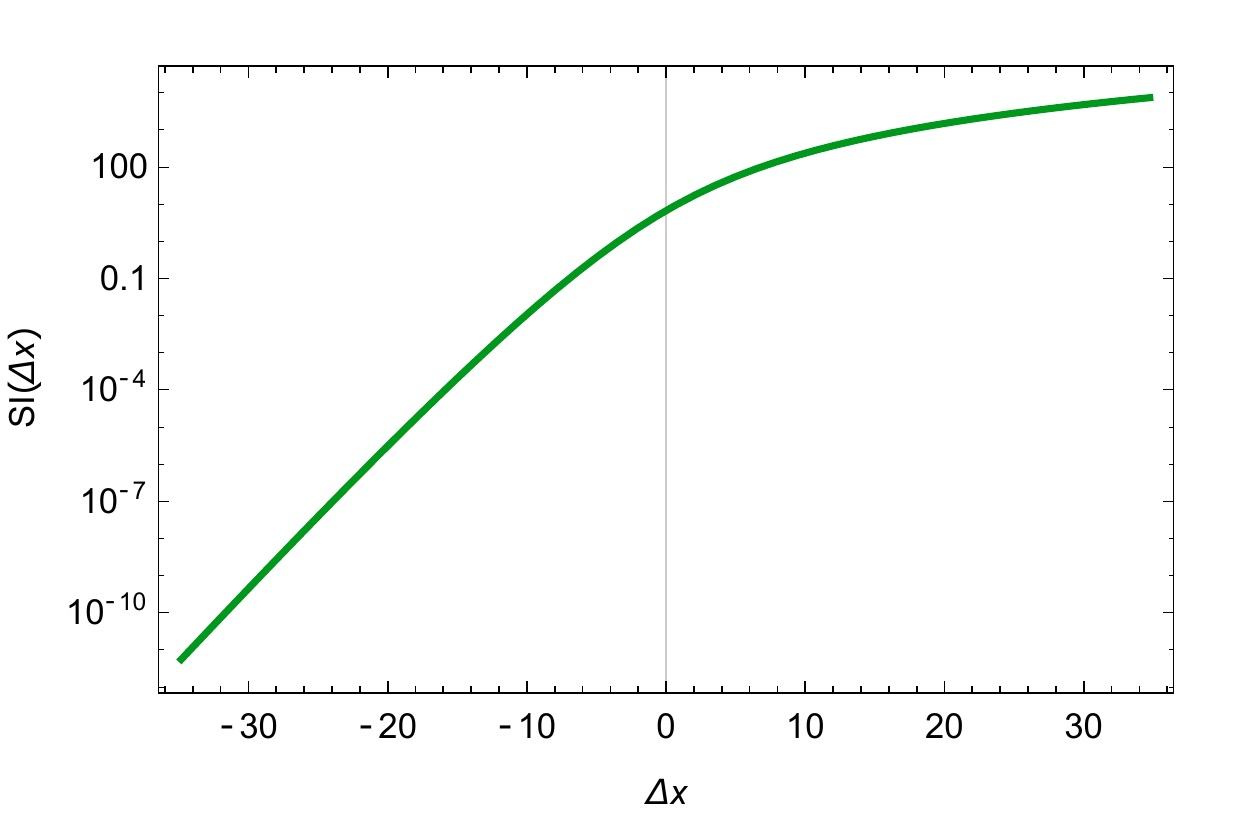}
\caption{\label{fig:sup5} The scattering integral $\mbox{SI}(\Delta x)$ as a function of $\Delta x $ plotted on a logarithmic scale. } 
\end{figure}

\noindent which we calculated numerically. $\mbox{SI}(\Delta x)$ is shown in Fig. \ref{fig:sup5}. We stress the property $\mbox{SI}(\Delta x)/\mbox{SI}(-\Delta x) = \mbox{exp}(\Delta x)  $, which guarantees the fixed equilibrium to fulfill Boltzmann statistics, i.e., in equilibrium the average occupation of level $s$ in subsystem $i$ satisfies $f_{i,s\pm 1}=f_{i,s}\mbox{exp}(\mp\Delta_i/k_B T_e)$. The scattering rate constant $\eta_{ij}$ is given by 

\begin{equation}
\label{eq:eta}
\eta_{ij} = \dfrac{2\pi}{\hbar} j_{\mathrm{ex},ij}^2 D_F^4   D_{\mathrm{s},i} D_{\mathrm{s},j} k_B^3,
\end{equation}

\noindent where $D_F$ is the free electron density of states (in units $\mbox{eV}^{-1}$). Now we define 

\begin{eqnarray}
W^{\pm \mp}_{ij,ss'} (\Delta_{ij})  &=& \mbox{SI}(\mp\Delta_{ij} /k_B T_e)  S_{ij,ss'}^{\pm\mp},
\end{eqnarray}

\noindent which corresponds to the dimensionless part of $\Gamma^{\pm\mp}_{ij, s s'}$. The time derivative of the average occupation $f_{i,s}$ of level $s$ is given by

\begin{eqnarray}
\dfrac{df_{i,s}}{dt} \bigg|_{\mbox{ex}} &=& \dfrac{\eta_{ij} C_j T_e^3}{D_{\mathrm{s},i}} \Big[ -f_{i,s} \sum^{S_j}_{s'=-S_j+1} W^{+-}_{ij,ss'} (\Delta_{ij}) f_{j,s'} 
-f_{i,s} \sum^{S_j-1}_{s'=-S_j} W^{-+}_{ij,ss'} (\Delta_{ij}) f_{j,s'} 
\\
\nonumber
&& \qquad \qquad   +f_{i,s-1} \sum^{S_j}_{s'=-S_j+1} W^{+-}_{ij,s-1 s'} (\Delta_{ij}) f_{j,s'} 
+f_{i,s+1} \sum^{S_j-1}_{s'=-S_j} W^{-+}_{ij,s+1 s'} (\Delta_{ij}) f_{j,s'}\Big] .
\end{eqnarray}

\noindent An expression for the normalized magnetization can by found by using $m_i = (-1/S_i)\sum_{s=-S_i}^{S_i} s f_{i,s} $. We find

\begin{eqnarray}
\label{eq:fermi} 
\dfrac{dm_i}{dt}\bigg|_{\mbox{ex}} &=& \dfrac{2 \eta_{ij} C_j}{\mu_{\mathrm{at},i}} T_e^3
 \bigg[
 \sum_{s=-S_i+1\,\,}^{S_i } \sum_{s=-S_j}^{S_j-1}  W^{-+}_{ij, ss'}(\Delta_{ij} )f_{i,s} f_{j,s'}-\sum_{s=-S_i\,\,}^{S_i-1} \sum_{s=-S_j+1}^{S_j } W^{+-}_{ij, ss'}(\Delta_{ij} )f_{i,s} f_{j,s'}  
 \bigg],
\end{eqnarray}

\noindent where we used $D_{\mathrm{s},i}=\mu_{\mathrm{at},i}/2S_i$. For $S_i=S_j=1/2$ this reduces to 

\begin{eqnarray}
\dfrac{dm_i}{dt}\bigg|_{\mbox{ex}} &=& \dfrac{2\eta_{ij} C_j}{\mu_{\mathrm{at},i} } T_e^3
 \bigg[   \mbox{SI}\bigg(\dfrac{\Delta_{ij}}{k_B T_e} \bigg)\dfrac{1-m_i}{2}\dfrac{1+m_j}{2}  -\mbox{SI}\bigg(-\dfrac{\Delta_{ij}}{k_B T_e} \bigg) \dfrac{1+m_i}{2}\dfrac{1-m_j}{2}
 \bigg].
\end{eqnarray}

\noindent Finally, the full  dynamics of spin subsystem $i$ is given by

\begin{eqnarray}
\dfrac{dm_i}{dt} = \dfrac{dm_i}{dt}\Big|_{\mathrm{EY}} + \dfrac{dm_i}{dt}\Big|_{\mbox{ex}}.
\end{eqnarray}

\clearpage

\section{III.$\qquad$The exchange scattering rate constant }

\noindent Although the coupling strength of the exchange scattering $j_{\mathrm{ex},ij}$ and the exchange coupling constant $j_{ij}$ have the same origin, the two can not be directly related  \cite{Schellekens2013}. We define the dimensionless free parameter $\lambda_{ij}$ that parametrizes the exact ratio between the coupling constants. Thus, we write 

\begin{equation}
\eta_{ij} = \dfrac{2\pi}{\hbar} j_{\mathrm{ex},ij}^2 D_F^4   D_{\mathrm{s},i} D_{\mathrm{s},j} k_B^3
\equiv \lambda_{ij} \dfrac{2\pi}{\hbar} j_{ij}^2 D_F^4   D_{\mathrm{s},i} D_{\mathrm{s},j} k_B^3.
\end{equation}

\noindent In the main text we assume that the ratios between the specific exchange scattering rate constants $\eta_{ij}$ are fully determined by the ratio between the exchange coupling constants $j_{ij}$. Hence, we take $\lambda_{\mathrm{Co}\mbox{-}\mathrm{Co}}=\lambda_{\mathrm{Co}\mbox{-}\mathrm{Gd}}=\lambda_{\mathrm{Gd}\mbox{-}\mathrm{Gd}}=\lambda$. Then we take $\lambda=5$ to retrieve realistic results, e.g., according to the experiments the concentration range where switching is found should be $\sim6 \% $ wide and the minimum threshold should be close to the compensation point \cite{Xu2017}. The value $\lambda=5$ sufficiently meets these requirements,  as can be seen in Fig. 2(a) of the main text. For comparison, the phase diagrams for different values of $\lambda$ are presented in Fig. \ref{fig:sup2} and \ref{fig:sup3}.

\section{IV.$\qquad$Used system parameters}

\noindent For simplicity, we assume that the electronic and phononic properties of the system are fully determined by the majority compound, i.e. Co, and described by a two-temperature model

\begin{eqnarray}
\gamma T_e(t) \dfrac{dT_e(t) }{dt} &=& g_{ep} (T_p(t) - T_e(t)) +P(t) ,
\\
C_p \dfrac{dT_p}{dt} &=& g_{ep} (T_e(t)- T_p(t)) +C_p \dfrac{T_{\mathrm{amb}} - T_p(t) }{\tau_D  }. 
\end{eqnarray}

\noindent 
where $T_{\mathrm{amb}}$ is the ambient temperature and $\tau_D$ is the time scale of the heat diffusion. For the simulations we set $T_{\mathrm{amb}}=295\mbox{ K}$ and $\tau_D=20\mbox{ ps}$. The other parameters for the electronic and phononic system are listed in table \ref{tab:table1}.

For the magnetic parameters of the system, e.g. $j_{ij}$ and $\gamma_{ij}$ we take the following approximations. The intra-sublattice exchange coupling should be chosen in such a way that for pure Co or pure Gd the correct Curie temperature is retrieved from the Weiss model. Hence, we should have that 

\begin{eqnarray}
\gamma_{\mathrm{Co}\mbox{-}\mathrm{Co}} &=& \dfrac{3 k_B T_{C,\mathrm{Co}}}{ S_{\mathrm{Co}}+1}  ,
\\
\gamma_{\mathrm{Gd}\mbox{-}\mathrm{Gd}} &=& \dfrac{3 k_B T_{C,\mathrm{Gd}}}{ S_{\mathrm{Gd}}+1}  .
\end{eqnarray}

\noindent The intra-sublattice exchange coupling constants can then be found by using that $\gamma_{ij} = z j_{ij} D_{s,j} S_j$ . For $z=12$ this leads to 

\begin{eqnarray}
j_{\mathrm{Co}\mbox{-}\mathrm{Co}} &=& \dfrac{3 k_B T_{\mathrm{C,\mathrm{Co}}}}{12(S_{\mathrm{Co}}+1) (\mu_{\mathrm{\mathrm{at,Co}}}/2) },
\\
j_{\mathrm{Gd}\mbox{-}\mathrm{Gd}} &=& \dfrac{3 k_B T_{C,\mathrm{Gd}}}{12(S_{\mathrm{Gd}}+1) (\mu_{\mathrm{at,Gd}}/2) }.
\end{eqnarray}

\noindent For the intersublattice exchange coupling constant $j_{\mathrm{Co}\mbox{-}\mathrm{Gd}}$, we assume that the ratio between $j_{\mathrm{Co}\mbox{-}\mathrm{Co}}$ and $j_{\mathrm{Co}\mbox{-}\mathrm{Gd}}$ is equal to the ratio used for Fe and Gd reported in \cite{Gerlach2017}. This means that we use $j_{\mathrm{Co}\mbox{-}\mathrm{Co}}\mu_{\mathrm{at,Co}}^2 : j_{\mathrm{Co}\mbox{-}\mathrm{Gd}} \mu_{\mathrm{at,Co}} \mu_{\mathrm{at,Gd}} \sim -0.388 $. Hence, we write

\begin{eqnarray}
j_{\mathrm{Co}\mbox{-}\mathrm{Gd}} &=& -0.388 \times \dfrac{3 k_B T_{C,\mathrm{Co}}}{12(S_{\mathrm{Co}}+1) (\mu_{\mathrm{at,Gd}}/2)}.
\end{eqnarray} 

\noindent From this, $\gamma_{\mathrm{Co}\mbox{-}\mathrm{Gd}}$ and $\gamma_{\mathrm{Gd}\mbox{-}\mathrm{Co}}$ can be easily calculated by using the definition $\gamma_{ij} = z j_{ij} D_{s,j} S_j$. Note that $\gamma_{\mathrm{Co}\mbox{-}\mathrm{Gd}}/\gamma_{\mathrm{Gd}\mbox{-}\mathrm{Co}} = \mu_{\mathrm{at,Gd}} /\mu_{\mathrm{at,Co}} $. This asymmetry is found to be important in  how small the concentration range for switching the alloys is. The used magnetic system parameters are listed in table \ref{tab:table2}.

\begin{table}[h!]
\caption{\label{tab:table1}%
Used parameters for the electron and phonon system \cite{Koopmans2010}. 
}
\begin{ruledtabular}
\begin{tabular}{llll}
$D_F$ [$\mbox{eV}^{-1} \mbox{at}^{-1}$] & $C_p$ [$\mbox{J }\mbox{m}^{-3}\mbox{K}^{-1}$] & $g_{ep}$ [$\mbox{J }\mbox{m}^{-3}\mbox{K}^{-1}\mbox{ps}^{-1}$] & $\gamma $ [$\mbox{J }\mbox{m}^{-3}\mbox{K}^{-2}$] \\
\colrule
 3 &  $4\cdot 10^6$ & $4.05\cdot 10^6$ & $2.0\cdot 10^3$
\end{tabular}
\end{ruledtabular}
\end{table}

\begin{table}[h!]
\caption{\label{tab:table2}%
Magnetic parameters used in the calculations \cite{Koopmans2010,Coey2009}.
}
\begin{ruledtabular}
\begin{tabular}{lllll}
\textrm{}&
$R$ [$\mbox{ps}^{-1}$] &
$T_C$ [$\mbox{K}$] & 
$\mu_{\mathrm{at}}$ [$\mu_B$] &S \\
\colrule
\mbox{Co} & 25.3 & 1388 & 1.72 & 1/2 \\
\mbox{Gd} & 0.092 & 292 & 7.55 & 7/2 \\
\end{tabular}
\end{ruledtabular}
\end{table}

\section{V.$\qquad$Notes on the inset of Figure 3 }

\noindent 

\begin{figure}[h!]
\includegraphics[scale=0.98]{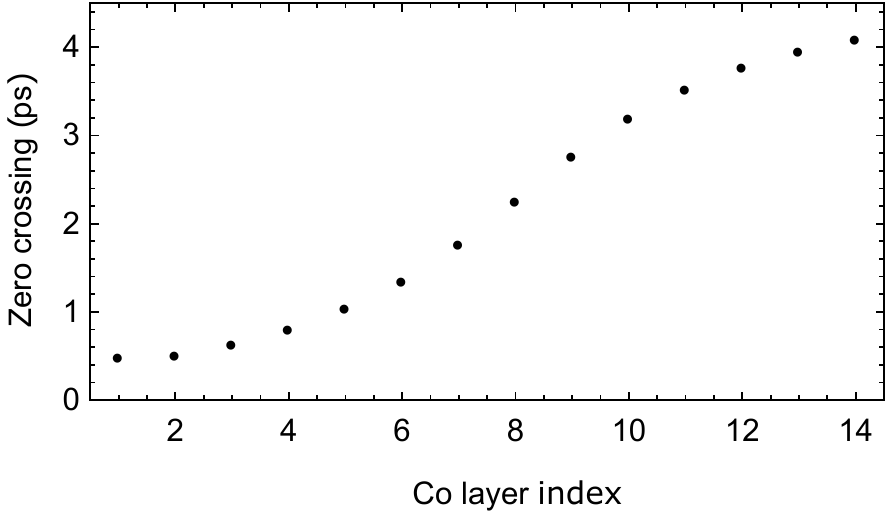}
\caption{\label{fig:sup11} The time at which each Co monolayer is reversed (zero crossings) as a function of the Co layer index, in a system of 14 Co monolayers and 3 Gd monolayers. The figure shows the same situation as the inset of Figure 3 in the main text. However, here we used $\lambda_{\mathrm{Co}\mbox{-}\mathrm{Gd}}=\lambda_{\mathrm{Gd}\mbox{-}\mathrm{Gd}}=5$ and $\lambda_{\mathrm{Co}\mbox{-}\mathrm{Co}}=1$. 
} 
\end{figure}

In the main article Fig. 3 (inset), we observed that the Co monolayer adjacent to the interface shows unexpected behaviour and does not switch first. Here, we discuss some extra details about this unintuitive effect.  In our theory we only include nearest neighbour interactions. Hence, only the Co monolayer adjacent to the interface experiences an exchange field originating from the Gd layer. The exchange field opposes the demagnetization of the inner Co monolayer, eventually leading to a delayed demagnetization and switching of that Co monolayer. The effect is strongly affected by the exchange scattering rate in the Co layer. Figure \ref{fig:sup11} shows the time at which each separate Co monolayer is switched in a system of 14 Co monolayers and 3 Gd monolayers after laser pulse excitation at $t=0$. It corresponds to the same system as the inset of Fig. 3 in the main text. In contrast to the main text, the calculation presented here corresponds to  $\lambda_{\mathrm{Co}\mbox{-}\mathrm{Gd}}=\lambda_{\mathrm{Gd}\mbox{-}\mathrm{Gd}} = 5$ and $\lambda_{\mathrm{Co}\mbox{-}\mathrm{Co}}=1$, i.e., we decreased the used value for $\eta_{\mathrm{Co}\mbox{-}\mathrm{Co}}$. Here, we clearly see that now the Co layer adjacent to the interface switches first. 

Finally, comparing Fig. \ref{fig:sup11} ($\lambda_{\mathrm{Co}\mbox{-}\mathrm{Co}}=1$) with    Fig. 3 from the main paper ($\lambda_{\mathrm{Co}\mbox{-}\mathrm{Co}}=5$), clearly shows that increasing the exchange scattering rate $\eta_{\mathrm{Co}\mbox{-}\mathrm{Co}}$ leads to a faster switching mechanism ($\sim 2$ times as fast). This again emphasizes that the exchange scattering between adjacent Co monolayers drives the switch throughout the Co layer.

\clearpage 

\section{VI.$\qquad$ Phase diagrams for different exchange scattering rates} 

\begin{figure}[t!]
\includegraphics[scale=0.96]{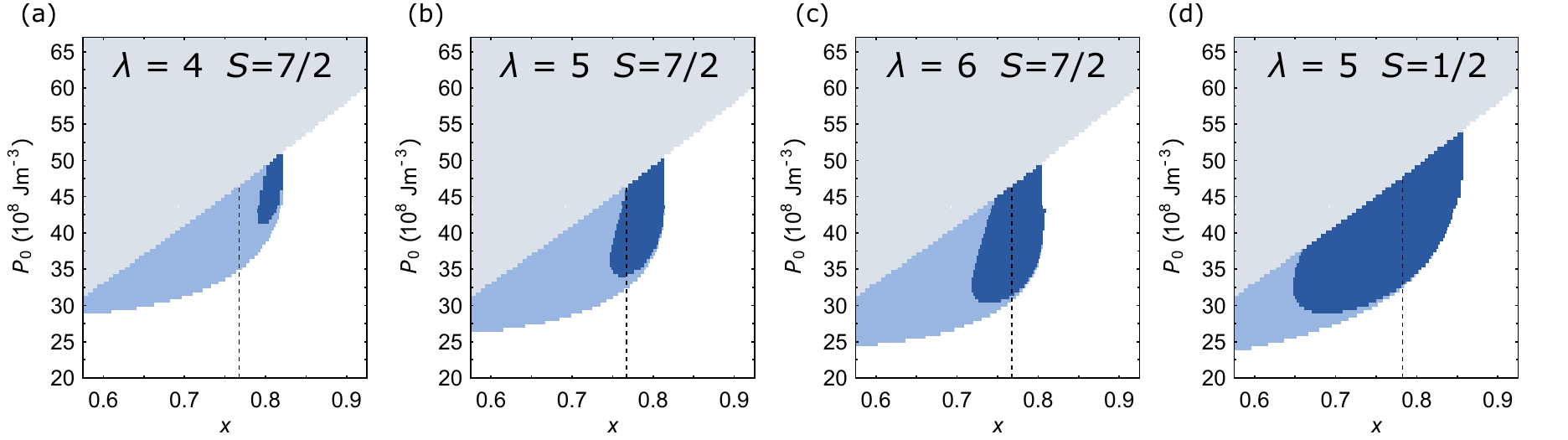}
\caption{\label{fig:sup2} Switching phase diagrams for $\mbox{Gd}_{1-x}\mbox{Co}_x$ alloys as a function of the Co concentration $x$ and the laser power $P_0$. Different values for $\lambda$ and spin quantum number $S_{\mathrm{Gd}}$ are used, which are given in the figures (a)-(d). The dashed lines indicate the compensation point $x_{\mathrm{comp}} $. 
} 
\end{figure}

\noindent Figure \ref{fig:sup2} shows the phase diagrams for AOS in $\mbox{Gd}_{1-x}\mbox{Co}_x$ as a function of the Co concentration $x$ and laser power $P_0$, for different values of $\lambda$ and $S_{\mathrm{Gd}}$ ($S_{\mathrm{Co}}=1/2$ for all figures). The figure shows that the model for the alloys is very sensitive to the choice of $\lambda$. Furthermore, the center, the width and the minimum of the dark blue region shifts for different choices of $\lambda$. Figure \ref{fig:sup2}(b) is in closest correspondence with the experimental findings \cite{Xu2017}, thereby  the value $\lambda=5$ is used in the main text. Note that setting $S_\mathrm{Gd}=7/2$ gives significantly better results compared to $S_{\mathrm{Gd}}=1/2$, e.g., for the latter the dark blue region is much too wide compared to the experiments (even for smaller values of $\lambda$). This is expected from the fact that the magnetization as a function of temperature is well described by a Weiss model with $S_\mathrm{Gd}=7/2$ \cite{Coey2009}.  

\begin{figure}[t!]
\includegraphics[scale=0.96]{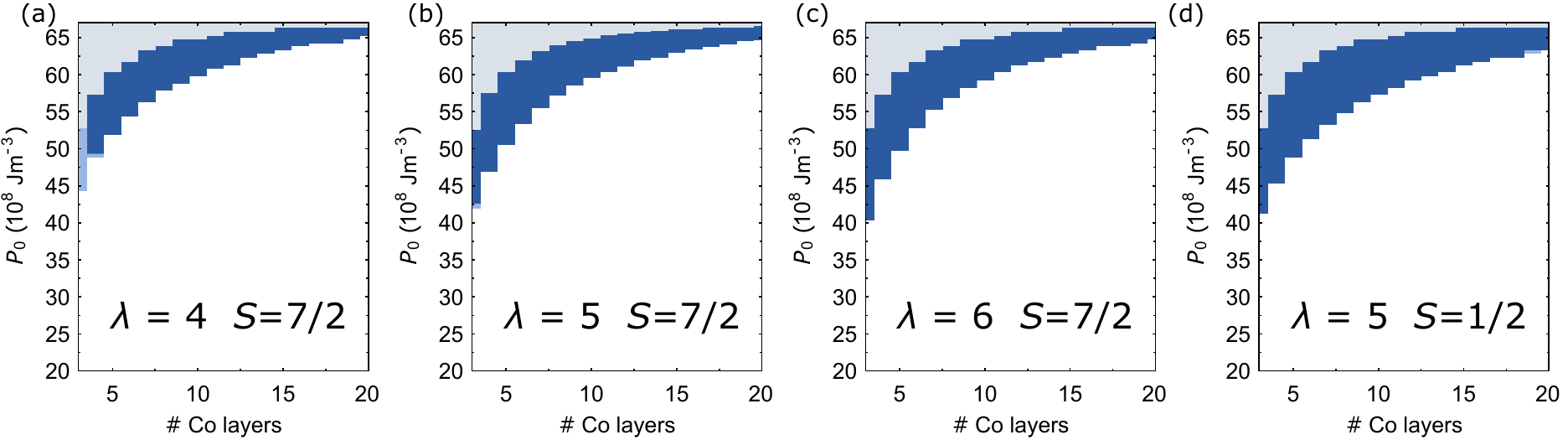}
\caption{\label{fig:sup3} Switching phase diagrams of Co/Gd bilayers as a function of the number of  Co monolayers and the laser power $P_0$. Different values for $\lambda$ and spin quantum number $S_{\mathrm{Gd}}$ are used, which are given in the figures (a)-(d).
} 
\end{figure}

Figure \ref{fig:sup3} clearly shows that the model for the bilayers is less sensitive to changes of $\lambda$ and $S_{\mathrm{Gd}}$. The reason for this is that the form of the switching (dark blue) region is mainly determined by the fact that $\eta_{\mathrm{Co}\mbox{-}\mathrm{Co}}\gg \eta_{\mathrm{Co}\mbox{-}\mathrm{Gd}}, \eta_{\mathrm{Gd}\mbox{-}\mathrm{Gd}}$. Even if $\lambda$ is decreased, $\eta_{\mathrm{Co}\mbox{-}\mathrm{Co}}$ still has a relatively large value, which corresponds to an efficient propagation mechanism. However, when $\lambda$ is decreased further (e.g. $\lambda=1$), no switching is found at all, because $\eta_{\mathrm{Co}\mbox{-}\mathrm{Gd}}$ and $\eta_{\mathrm{Gd}\mbox{-}\mathrm{Gd}}$ need to be sufficiently large  to switch the layers near the interface (and to nucleate the front of reversed Co magnetization at the interface). In conclusion, when $\lambda$ is chosen such that switching is found in the model, the propagation mechanism is always present due to the relatively large Co-Co scattering rate $\eta_{\mathrm{Co}\mbox{-}\mathrm{Co}}$. Hence, the qualitative statements in the main text about the propagation mechanism in the Co/Gd bilayers, are independent of the exact choice of $\lambda$.

\clearpage 

\section{VII.$\qquad$Experimental domain size as a function of pulse energy}

\begin{figure}[h!]
\includegraphics[scale=0.35]{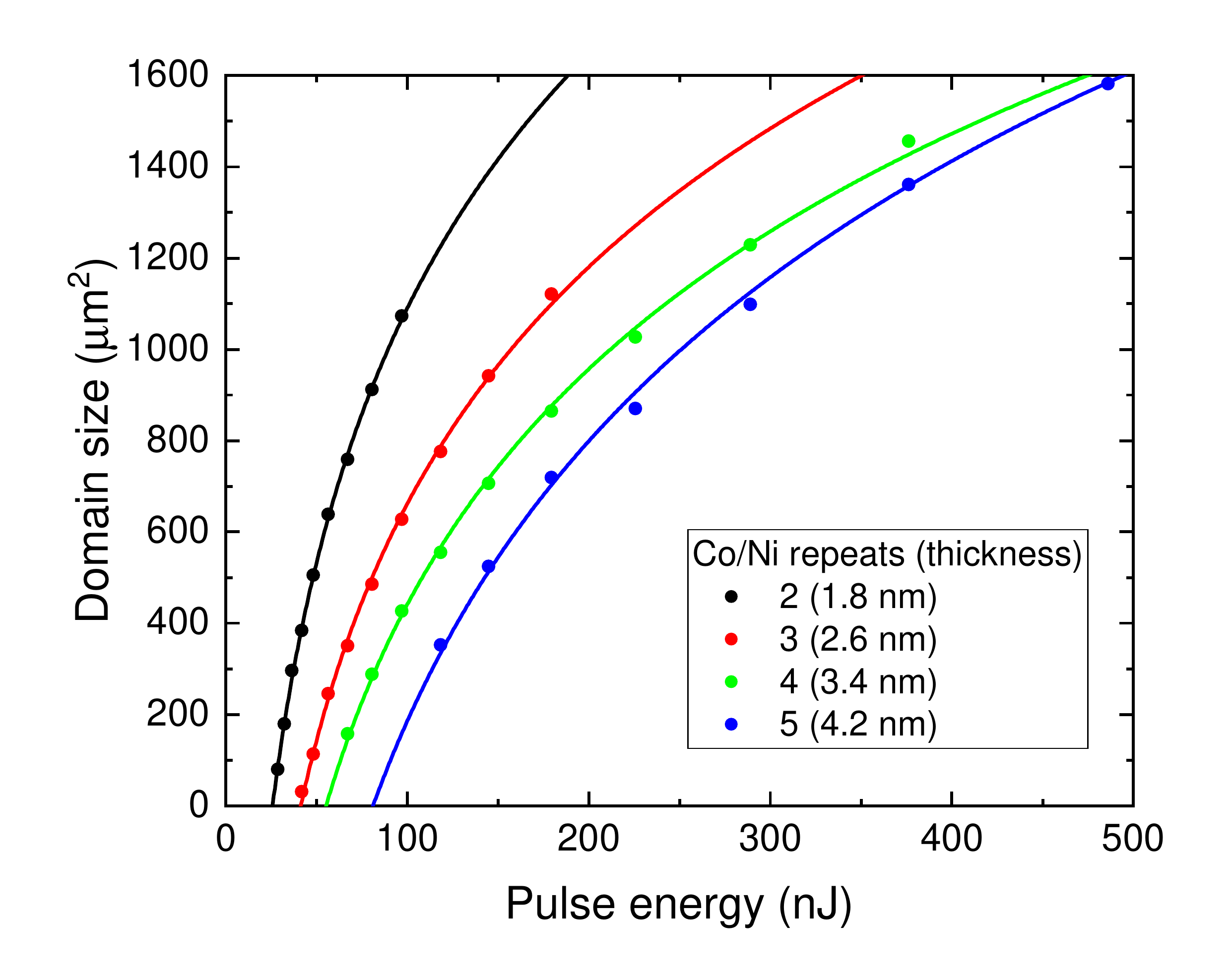}
\caption{\label{fig:sup4} Domain size as a function of pulse energy for Pt/Co/[Ni/Co]$_N$ /Gd stacks with N = 2, 3, 4, and 5.
} 
\end{figure}

\noindent Figure \ref{fig:sup4} shows a plot of the experimental domain size as a function of the pulse energy for the Pt/Co/[Ni/Co]$_N$/Gd stacks with $N=2,3,4$ and $5$. The fit functions can be used to determine the threshold fluence (Fig. 1 in the main paper). For the exact method we refer to \cite{Lalieu2017} and \cite{Liu1982}.  

\providecommand{\noopsort}[1]{}\providecommand{\singleletter}[1]{#1}%